# VCF2Networks: applying Genotype Networks to Single Nucleotide Variants data

Giovanni Marco Dall'Olio[1], Ali R. Vahdati[2], Bertranpetit Jaume[1], Wagner Andreas[2,3,4], Laayouni Hafid[1]

[1] Institut de Biologia Evolutiva (CSIC-Universitat Pompeu Fabra), 08003 Barcelona, Catalonia, Spain

[2] Institute of Evolutionary Biology and Environmental Studies, University of Zurich.

[3] The Swiss Institute of Bioinformatics, Bioinformatics, Quartier Sorge, Batiment Genopode, 1015 Lausanne, Switzerland.

[4] The Santa Fe Institute, 1399 Hyde Park Road, Santa Fe, NM 87501, USA

## Abstract

**Summary**: Genotype networks are a method used in systems biology to study the innovability of a given phenotype, determining whether the phenotype is robust to mutations, and how do the genotypes associated to it are distributed in the genotype space. Here we developed VCF2Networks, a tool to apply this method to population genetics data, and in particular to single Nucleotide Variants data encoded in the Variant Call file Format (VCF).
A complete summary of the properties of the genotype network that can be calculated by VCF2Networks is given in the Supplementary Materials 1.

**Availability and Implementation**: The home page of the project is https://bitbucket.org/dalloliogm/vcf2networks . VCF2Networks is also available directly from the Python Package Index (PyPI), under the name vcf2networks.

## Introduction

Genotype Networks are a method used in systems biology to study the innovability of a given phenotype. It is implemented by building a network representation of all the genotypes associated with the phenotype, and calculating some properties of the network that, in the literature, have been shown to be associated with innovability. Genotype Networks can be used to infer whether the phenotype is robust to mutations, or whether it shows high genetic heterogeneity (Wagner 2011). They have been applied to a wide range of systems, like genetic circuits (Wagner, 2003; Espinosa-Soto *et al.*, 2011; Ciliberti *et al.*, 2007), metabolic networks (Wagner, 2009, 2007; Matias Rodrigues and Wagner, 2009; Samal *et al.*, 2010; Dhar *et al.*, 2011), and RNA and protein foldings (Ferrada and Wagner, 2010; Schultes and Bartel, 2000). In recent times, they have also been used to reconcile the neutralist and selectionist views on evolution (Wagner, 2008a), and to examine exaptation in metabolic networks (Barve and Wagner, 2013).

So far, the application of genotype networks to genomics data has been limited, because the number of samples needed to construct such networks is high (Dall'Olio et al, 2013). In fact, genotype networks can be calculated efficiently only when large datasets of genotypes are available. Fortunately, thanks to the cheapening of genome sequencing, this difficulty is being alleviated. This makes the application of the concept of genotype networks to population data a more likely possibility, allowing to broadening our understanding of the genetic variation behind a phenotype.

## Approach and purpose

Here we present VCF2Networks, a script to calculate many properties of the genotype network of a region, from a Variant Call Format (VCF) (Danecek *et al.*, 2011) file.

In our implementation, a genotype network is a network in which all the nodes are strings representing genotypes associated with the phenotype, and in which edges connect nodes that have only one single difference. As an example, Figure 1 in Supplementary Materials I shows the genotype network of a region comprising 5 Single Nucleotide Variants (SNVs), in a hypothetical population (only the green nodes belong to the genotype network, the others are shown for comparison). In the network, genotypes are encoded as strings of “0”s and “1”s, where the “0”s represent the reference allele, and the “1”s the alternative allele.

In the literature, properties such as the average path length and the average degree of the genotype network of a system have been associated with higher evolvability, ability to explore larger portion of the space, and with higher robustness to mutations (Hahn *et al.*, 2004; Wagner, 2008b; Payne *et al.*, 2013; Manrubia and Cuesta, 2010). Other properties of the network, such as the betweenness, or the average closeness of a network may give other insights on genome variation, once properly studied. A list of all the properties of the genotype networks that can be calculated with VCF2Networks is given in Supplementary Materials 1, along with a short description of how they can be interpreted in biological terms. A more complete review on genotype networks is provided in (Wagner, 2011).

## Usage

The installation process will produce an executable called vcf2networks. The basic usage is the following:

$: vcf2networks --vcf mydata.vcf --individuals individuals_annotations.txt --phenotype disease_status

This command will parse the genotype data from the file mydata.vcf (--vcf option), the association of individuals with phenotypes from the file individuals_annotations.txt (--individuals option), and group individuals according to the phenotype "disease_status" (--phenotype option). The output will be a tabular report as in Table 1:

| file | disease_status | n_datapoints | n_components | av_path_length | av_degree | window | snps |
|---|---|---|---|---|---|---|---|
| mydata | glob | 2184 | 17 | 3.8470 | 2.5037 | window_0 | 37 |
| mydata | disease1 | 758 | 7 | 2.6508 | 2.3256 | window_0 | 37 |
| mydata | disease2 | 572 | 6 | 2.9000 | 2.0000 | window_0 | 37 |
| mydata | control | 362 | 8 | 2.6477 | 2.1277 | window_0 | 37 |

*Table 1: example of output from vcf2networks*

The example output in Table 1 contains the Number of Components, Average Path Length, and Average Degree of the genotype networks built from the mydata.vcf file, among with other descriptive columns. The first row shows the values for the "glob" group, corresponding to all the individuals in the VCF file (as for "global" population). The remaining columns show the values for each group of individuals, as defined in the individuals_annotations.txt file.

The documentation in the VCF2Networks website provides many more examples of usage and options. Notably, the script allows to group individuals by any given phenotype (e.g. disease status, eye color, etc..), as long as the phenotype is defined in the individuals_annotations.txt file. It is also possible to calculate other properties of the network, such as the Diameter, the Average Closeness, or the Maximum Betweenness (see Supplementary Materials 1). Finally, it is possible to subsample only a given number of genotypes from each population, to apply a sliding window approach, or to customize the criteria used to define whether two nodes are neighbor in the genotype network.

## Availability

VCF2Networks is available from the Python Package Index, under the name vcf2networks. The recommended way to install it is through the python setuptools utilities, or any equivalent:

```
$: easy_install vcf2networks
```

The home page of the project is https://bitbucket.org/dalloliogm/vcf2networks/ . The implementation follows the rules suggested in (Seemann, 2013).

## Acknowledgements

This research as funded by grants BFU2010-19443 (subprogram BMC) awarded by Ministerio de Ciencia y Tecnología (Spain) and by the Direcció General de Recerca, Generalitat de Catalunya (Grup de Recerca Consolidat 2009 SGR 1101). GMD is supported by a FPI fellowship BES-2009-017731. We thank Tiago Carvalho, Brandon Invergo, Juan Ramón González Vallinas, and Christian Pérez-Llamas for helping testing the software.

## Supplementary Materials 1: Description of Network Properties

This file contains a description of all the possible columns of the output of vcf2networks. Some of these columns describe the data used to generate the network (e.g. the position and size of the region, the id of the central SNV, etc..), while other columns describe the network properties that can be calculated (e.g. the average degree, the diameter of the network, etc..). To customize the columns included in the output file, use the --config option.

### 1. Network Definition Properties:

This first group of network properties describes the parameters used to generate the networks. This includes details like which individuals have been used (grouped in continents or phenotype groups), the definition of distance used, and the length of the window.

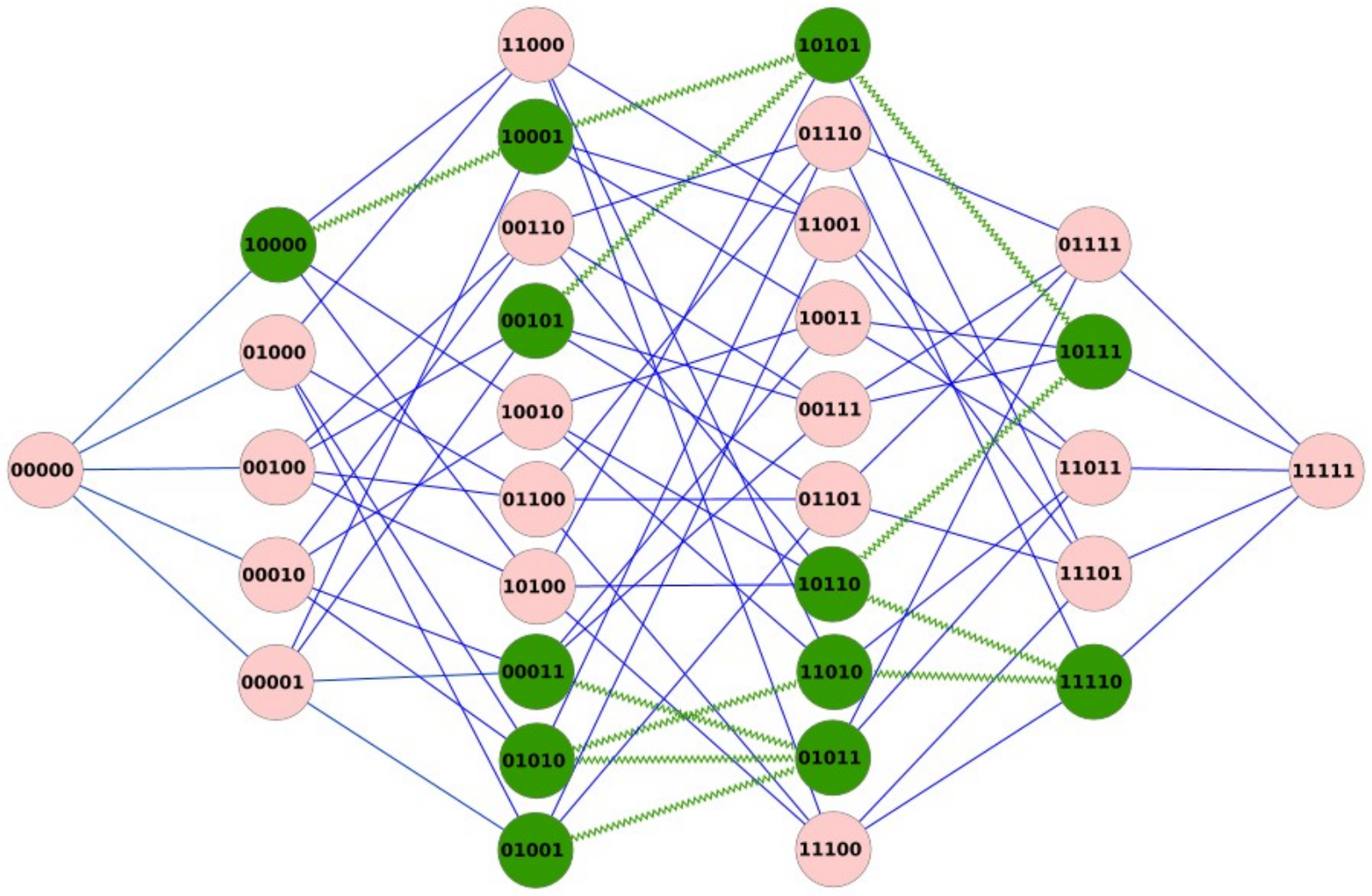

*figure 1: Definition of Genotype Network (green nodes), displayed in comparison with the whole genotype space (whole network, including pink and green nodes). Green nodes mark the genotypes that are present at least once in a population. Pink nodes are not present in the population, and do not belong to the genotype network, but are shown to better illustrate how the genotype network occupies the space of possible genotypes. For this network, distance_definition is 1 and n_SNVs is 5.*

- **file_prefix**: a reference to the file used to generate the network, without the suffix. This is useful when merging the results from multiple datasets.
- **continent**: name of the continent group, or the population, to which the individuals used to generate the network belong. If the --phenotype option is not provided, vcf2networks groups individuals by continental group, producing one output line for the global population (all the individuals in the VCF file), plus one line per every population.
- **phenotype**: if the --phenotype option is given, the second column of the output file will contain the phenotype group to which each network belongs. By default, *vcf2networks* will generate one line for the global population, plus one distinct line for each group defined in the phenotype column requested.

- **distance_definition**: the definition of distance between two nodes in a network. Two genotypes are considered to be connected if they have this number of differences or less. For example, the genotypes 0000 and 0011, which differ in two positions, are considered to be connected only if distance_definition is two or more. The default value is one, meaning that two nodes are connected if they share only one difference.
- **n_SNVs**: the number of SNVs in the window. This is also referred to as "window size". If no --windows_size option is provided, vcf2networks will calculate a network on all the SNVs included in the file, and n_SNVs will be equal to this number of SNVs. If the --windows_size option is provided, the n_SNVs column will describe how many SNVs have been used to generate the network.

## 2. Window Position Properties

This group of properties give information on the position of the SNVs used to generate the network in the genome. These include details like what is the position in the chromosome, what is the size, which is the SNV in the center of the window, etc.. This information can be used for later analysis, to compare them with other annotations on the genome.

To analyze large regions of the genome, it is recommended to run a sliding window approach, splitting the region in smaller chunks, and calculating a genotype network for each chunk (see the -w and -l options in *generate_network_report.py*). Figure 2 describes an example of how the sliding overlapping windows algorithm can be implemented.

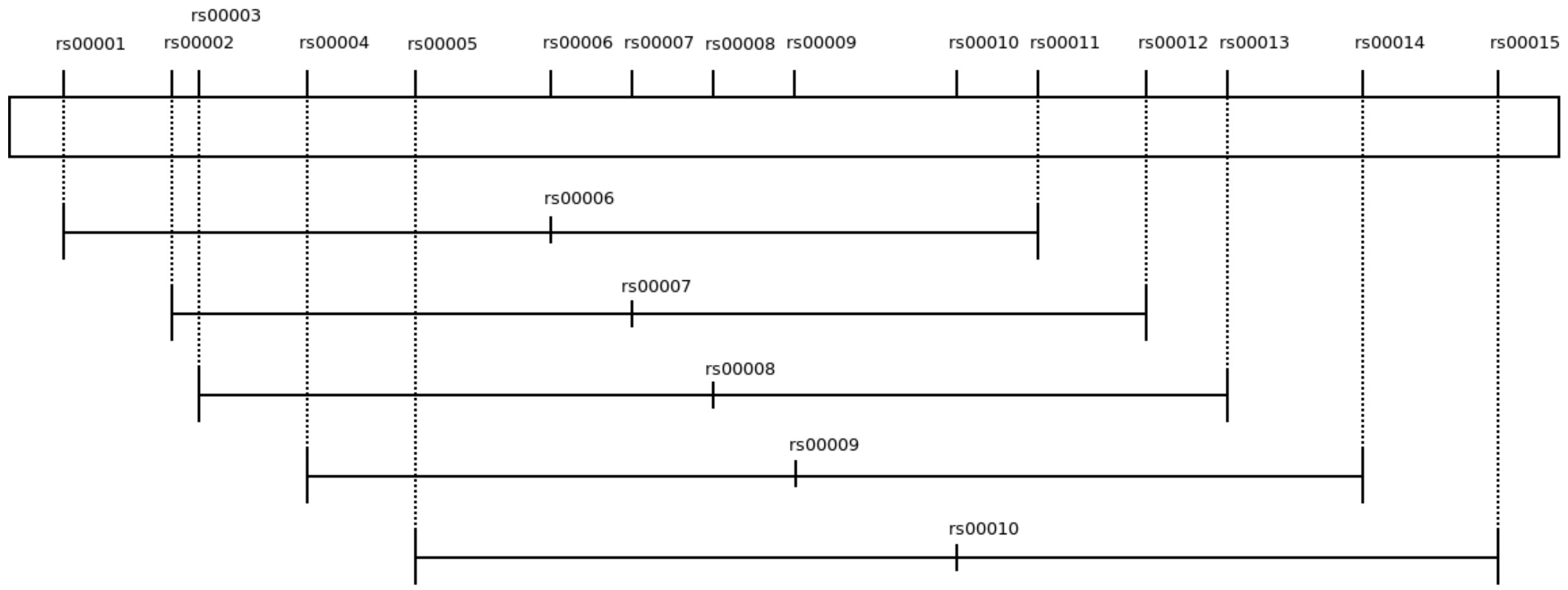

*figure 2: A scheme of how the sliding window approach can be applied. In this figure, the region is split in windows of 11 SNVs, by moving only 1 SNV at a time (options -w 11 -l in generate_network_report.py). In later analysis, the ID of the central SNV can be used as an unique key to refer to each window..*

- **central_SNV**: ID of the central SNV in the window, taken from the original VCF file. This value can be used as a unique key to refer to each network.
- **chromosome**: chromosome in which the SNVs used to generate the network are located.
- **central_SNV_position**: position of the central SNV of the network.
- **upstream_position**: position of the most upstream SNV of the network.
- **downstream_position**: position of the most downstream SNV of the network.
- **region_size**: size of the network in the genome, expressed in number of nucleotide bases.
- **distance_from_upstream_margin**: distance of the central SNV from the most upstream SNV used to generate the network.
- **distance_from_downstream_margin**: distance of the central SNV from the most downstream SNV used to generate the network.

## 3. Data Density Properties

This group contains information on the density of genotypes in the network. This includes details like how many individuals have a given genotype on average, and so on.

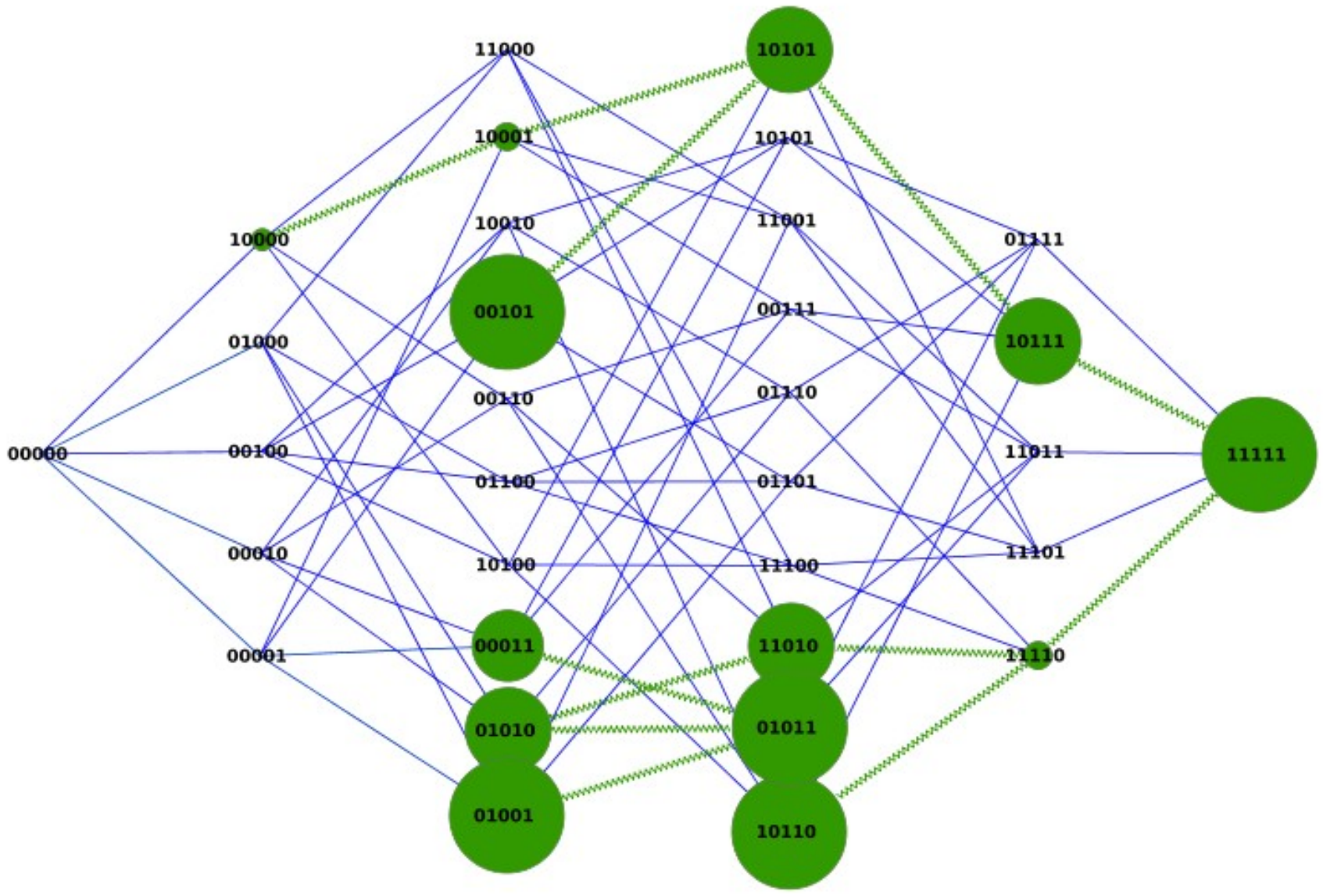

*figure 3: A Genotype Network, illustrating data density. Green nodes mark the genotypes that are present at least once in the population. Node size is proportional to the number of individuals that carry that genotype in the population.*

- **n_datapoints**: the number of samples (also known as haploid genotypes) used to build the network. For a diploid population, it is the double of the number of individuals in the dataset: for example, if there are 300 individuals, n_datapoints is 600. Certain options in *generate_network_report.py* allows to use only the individuals belonging to a population, or to randomly subsample a fixed number of samples.
- **av_datapoints_per_node**: how many samples there are for each node. For example, let's imagine that a network contains only two nodes, 000 and 001. The first, 000, is present in 200 samples, and the second, 001, is present in 400 samples. In this network av_datapoints_per_node would be 300. High values of av_datapoints_per_node means that most of the individuals have the same genotype.

### 4. Properties that explain how much a population has explored the genotype space

These properties describe how extensively a population has explored the genotype space. If there are no selective constraints, a population should be free to explore the space, reaching a high average path length and diameter. On the other hand, if there are functional or evolutionary constraints that inhibits the population from exploring the space, this will be reflected in a lower average path length and diameter.

- **n_vertices:** how many nodes (or vertices) there are in the network. This also corresponds to how many genotypes are present at least once in the population.
- **n_edges:** how many connections there are in the network. Two nodes are connected if they share less or exactly a given number of differences, defined in the distance_definition parameter.
- **n_components**: the number of connected components for each network. A connected component is a chain of nodes that are connected by at least one edge (see figure 5 below). This property describe how fragmented is the genotype space of a population. A large number of components means that there are a lot of gaps between individuals, which can be explained either by the presence of subpopulation structure, or by insufficient data.

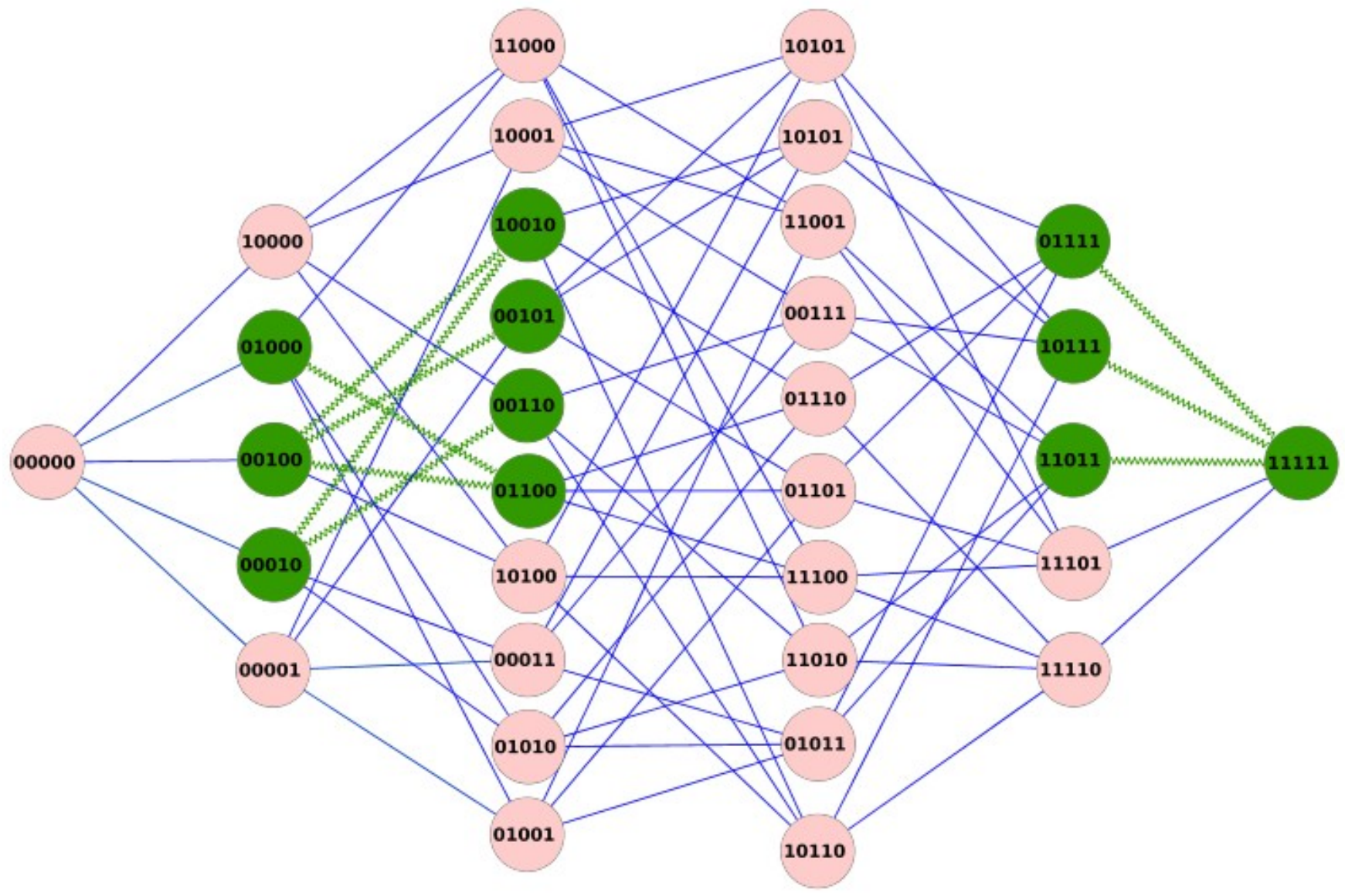

*figure 4: Connected Components. Green nodes represent the genotypes that are present at least once in a given population. In this case, some nodes are disconnected from the others*

- **av_path_length**: The average of all the paths in the graph. A large average path length means that the population has explored a vast portion of the genotype space.
- **var_path_length**: The variance of all path lengths in the graph. This can be used in combination with av_path_length to have an idea of how the path lengths are distributed.
- **diameter**: The longest path in the graph. It can be used in combination with the average path length, to determine how extensively a population has explored the space.
- **av_w_path_length**: The average of all paths in the graph, weighted by the data density of each

node. If two nodes are present in a lot of individuals, the presence of a path between them will be considered more important than the paths between nodes that are present in less individuals.

- **av_w_path_length_inv**: A complementary implementation of av_w_path_length. In this case, the weight of each node is the inverse of the weight used in av_w_path_length.
- **av_closeness**: The closeness is a measure of how “close” is a node to all the other nodes in the network. Thus, the average closeness of a networks measures how much the all the nodes of the network are close to each other. Low values of average closeness indicate that the nodes are close to each other, while high values of average closeness indicate that nodes are more distant.
- **var_closeness**: The variance of the distribution of all node closeness in the network.
- **median_closeness**: The median of the distribution of all node closeness in the network.
- **max_closeness**: The maximum of all node closeness in the network.

## 5. Properties that explain the robustness of a population to mutations

The robustness of a network describes how the network will respond to point mutations. In general, phenotypes whose genotype networks have a high average degree are considered to be more robust to mutations than others.

- **av_degree**: The average of all node degrees in the network. In principle, if a node has a high degree, it means that a mutation on it is more likely to end up in another node that is already in the network. So, the average degree can be used to identify if the network is robust to mutations.
- **var_degree**: The variance of all node degrees in the network. Useful to understand the distribution of node degrees.
- **median_degree**: The median of all node degrees in the network.
- **max_degree**: The maximum node degree in the network.
- **av_w_degree**: A weighted version of av_degree. In this case, the degree of each node is multiplied by the fraction of data points for that node. So, if a node corresponds to a genotype that is present in a lot of individuals, its contribution to the av_w_degree will be larger than other nodes.
- **av_w_degree_inv**: An alternative way to calculate the av_w_degree. In this case, the weights are the inverse of the weights used in av_w_degree.
- **density**: The network density measures how many connections there are in the network, compared to how many possible connections there it may be.

## 6. Properties that show the presence of clusters, and of bridge genotypes

The betweenness of a node is proportional to how many shortest paths pass through the node. The betweenness can be used to identify nodes that act as a bridge between two groups of genotypes.

- **av_betweenness**: The average of all node betweenness of the network. Usually, most nodes will have a low betweenness, so this average will not be much informative. The max betweenness is more useful.
- **var_betweenness**: The variance of all node betweenness in the network.
- **median_betweenness**: The median of all node betweenness in the network.
- **max_betweenness**: The maximum node betweenness in the network.

## 7. Other columns:

The output of vcf2networks will always contain the following columns, which are useful for reference purposes.

- **window**: a unique ID for each window, when the sliding window option is used.
- **whole_gene_nSNVs**: The total number of SNVs in the dataset, or in the chromosome analyzed.